\documentclass[aps,prd,preprint,a4paper,showpacs,nofootinbib,superscriptaddress]{revtex4-2}
\usepackage{bm}
\usepackage{indentfirst}
\usepackage{amsmath}
\usepackage{graphicx}
\usepackage{float}
\usepackage{amssymb}
\usepackage{subfigure}
\usepackage{amssymb}
\usepackage{hyperref}
\usepackage{epstopdf}
\usepackage[section]{placeins}

\usepackage[utf8]{inputenc}
\hypersetup{
    colorlinks=true,
    linkcolor=red,
    citecolor=blue,
}
\usepackage{color}
\usepackage[T1]{fontenc}
\usepackage{txfonts}
\usepackage{orcidlink}

\begin{document}

\title{Dynamic generation or removal of a scalar hair}

\author{Yunqi Liu}
\email{yunqiliu@yzu.edu.cn}
\address{\textit{Center for Gravitation and Cosmology, College of Physical Science and Technology, Yangzhou University, Yangzhou 225009, China}}
\address{\textit{School of Aeronautics and Astronautics, Shanghai Jiao Tong
University, Shanghai 200240, China}}

\author{Cheng-Yong Zhang}
\email{zhangcy@email.jnu.edu.cn (corresponding author)}
\address{\textit{Department of Physics and Siyuan Laboratory, Jinan University, Guangzhou 510632, China}}

\author{Wei-Liang Qian}
\email{wlqian@usp.br (corresponding author)}
\address{\textit{Escola de Engenharia de Lorena, Universidade de S\~ao Paulo, 12602-810, Lorena, SP, Brazil}}
\address{\textit{Faculdade de Engenharia de Guaratinguet\'a, Universidade Estadual Paulista, 12516-410, Guaratinguet\'a, SP, Brazil}}
\address{\textit{Center for Gravitation and Cosmology, College of Physical Science and Technology, Yangzhou University, Yangzhou 225009, China}}

\author{Kai Lin}
\email{lk314159@hotmail.com}
\address{\textit{Hubei Subsurface Multi-scale Imaging Key Laboratory, Institute of Geophysics and Geomatics, China University of Geosciences, 430074, Wuhan, Hubei, China}}
\address{\textit{Escola de Engenharia de Lorena, Universidade de S\~ao Paulo, 12602-810, Lorena, SP, Brazil}}

\author{Bin Wang}
\email{wang_b@sjtu.edu.cn}
\address{\textit{Center for Gravitation and Cosmology, College of Physical Science and Technology, Yangzhou University, Yangzhou 225009, China}}
\address{\textit{Shanghai Frontier Science Center for Gravitational Wave Detection,
Shanghai Jiao Tong University, Shanghai 200240, China}}
\baselineskip=0.5 cm
\begin{abstract}
We study dynamic processes through which the scalar hair of black holes is generated or detached in a theory with a scalar field non-minimally coupled to Gauss-Bonnet and Ricci scalar invariants.  
We concentrate on the nonlinear temporal evolution of a far-from-equilibrium gravitational system. 
In our simulations, we choose the initial spacetime to be either a bald Schwarzschild or a scalarized spherically symmetric black hole.
Succeeding continuous accretion of the scalar field onto the original black hole, the final fate of the system displays intriguing features, which depend on the initial configurations, strengths of the perturbation, and specific metric parameters.  
In addition to the scalarization process through which the bald black hole addresses scalar hair, we observe the dynamical descalarization, which removes scalar hair from an original hairy hole after continuous scalar field accretion. 
We examine the temporal evolution of the scalar field, the metrics, and the Misner-Sharp mass of the spacetime and exhibit rich phase structures through nonlinear dynamical processes.
\end{abstract}

\maketitle

\section{Introduction and main result}\label{intro}

One of the fascinating questions in General Relativity (GR) is the black hole's uniqueness theorem, also referred to as the {\it no-hair conjecture}~\cite{Carter:1971zc, Israel:1967wq, Ruffini:1971bza, Bekenstein:1995un, Sotiriou:2011dz}.
It states that the black hole solutions are entirely characterized by three quantities, namely, the mass, electric charge, and angular momentum.
The applicability of the no-hair conjecture has been extended to the Brans-Dicke theories, a few classes of scalar-tensor theories, and Gallilen models of gravity~\cite{Pani:2011gy, Herdeiro:2014goa, Babichev:2013cya}. 
In recent years, there has been a surge of interest in black hole solutions in the presence of a nonlinear field.
In particular, gravitational theories beyond GR or even GR with specific matter sources were shown to evade the no-hair conjecture and instead develop {\it hairy} black hole solutions featured by more than three quantities. 
Such cases include the black holes in the presence of the Yang-Mills~\cite{Volkov:1989fi, Bizon:1990sr, Greene:1992fw,Maeda:1993ap}, Skyrme~\cite{Luckock:1986tr,Droz:1991cx}, conformally-coupled scalar fields~\cite{Bekenstein:1974sf}, the dilatonic and colored black holes in the context of the Einstein-dilaton-Gauss-Bonnet theory (EdGB)~\cite{Torii:1996yi,Kanti:1996gs,Kleihaus:2015aje,Kleihaus:2011tg,Guo:2008hf}, as well as the rotating~\cite{Ayzenberg:2014aka,Maeda:2009uy,Ohta:2010ae} or shift-symmetric Galileon~\cite{Sotiriou:2014pfa,Benkel:2016rlz} hairy black holes.

Indeed, rather rich content regarding the phase diagram has been revealed in gravitational systems beyond GR~\cite{Cadoni:2009xm, Cardoso:2013fwa, Kleihaus:2015iea, Silva:2018qhn,  Antoniou:2021zoy}.
The latter is particularly interesting in theories potentially ``afflicted'' by the tachyonic instability~\cite{Damour:1993hw, Cardoso:2013opa, Cardoso:2013fwa, Zhang:2014kna}, that eventually gives rise to the {\it Spontaneous scalarization}.
Among others, the mechanism typically appears in the models containing a non-minimal coupling of a real scalar field to some source terms.
The source terms in question could be a geometrically invariant quantity such as the Gauss-Bonnet invariant in extended Scalar-Gauss-Bonnet theory~\cite{Doneva:2017bvd,Silva:2017uqg,Antoniou:2017acq,Cunha:2019dwb,Dima:2020yac,Herdeiro:2020wei,Berti:2020kgk,Corelli:2022pio,Corelli:2022phw}, the Ricci scalar for non-conformally invariant black holes~\cite{Herdeiro:2019yjy}, the Chern-Simons invariant~\cite{Brihaye:2018bgc}, and the Maxwell invariant $F_{\mu\nu}F^{\mu\nu}$ \cite{Herdeiro:2018wub}.
More recently, further developments have been carried out on the spontaneous scalarization associated with compact objects~\cite{Brihaye:2019puo,Astefanesei:2019pfq,Blazquez-Salcedo:2018jnn,Macedo:2019sem,Lin:2020asf,Guo:2020sdu,Bakopoulos:2020dfg}.
The presence of such non-minimal coupling plays the role of an effective mass in the equation of motion of the scalar field, and the tachyonic instability might be triggered when the mass becomes negative.

On the one hand, many studies to date, for the most part, focused on the static aspect, such as the linear (in)stability of a given (hairy) black hole spacetime.
On the other hand, the relevant processes that lead to the formation of black holes that evade the no-hair conjecture are dynamical ones.
Specifically, the temporal evolution is involved through which an unstable bald black hole solution in GR acquires a hair owing to the relevant scalar perturbations, which is, by definition, dynamic in nature.
Indeed, the relevant physical process involves the entire non-perturbative evolution from the initial state to the final one. 
Moreover, it is expected that the black hole hair can be either generated or deprived dynamically in terms of a nontrivial time-dependent scalar field configuration outside the horizon.
Although the linear stability properties of the hairy black holes have been extensively investigated~\cite{Zou:2019bpt, Myung:2020ctt, Silva:2018qhn, Hod:2019pmb, Konoplya:2019fpy, Blazquez-Salcedo:2020rhf, Zhang:2020pko}, studies regarding the temporal evolution associated with the scalarization furnish further information on the dynamical properties of the gravitational system that might be otherwise hidden from the static metric solution, such as the critical phenomena featured by discrete self-similarity uncovered at the threshold of the gravitational collapse \cite{Choptuik:1992jv, Liebling:1996dx}.
Moreover, the whereabouts of the endpoint of the instability can only be established through fully nonlinear numerical simulations. 
In particular, many intriguing results are not straightforward from the viewpoint of the linear stability analysis.
In this context, the nonlinear dynamics plays an essential role for gaining a deeper insight into the black hole scalarization and its reverse process, the so-called {\it descalarization}~\cite{Elley:2022ept, Doneva:2022byd, Zhang:2022cmu}.

In the case where the scalar is coupled to the electromagnetic field, known as the Einstein-Maxwell Scalar (EMS) theory, rich physics in the black hole scalarization and descalarization mechanisms were disclosed~\cite{Zhang:2022cmu,Zhang:2021etr, Zhang:2021edm,Zhang:2021nnn,Zhang:2021ybj}.
More recently, novel critical phenomena were revealed in the nonlinear dynamical processes involving accretion of scalar fields onto black holes~\cite{Zhang:2021etr, Zhang:2021edm}. 
The relevant spacetime configuration to ignite the scalarization was scrutinized, and the critical exponents associated with the scalarization and descalarization of the phase transitions were extracted~\cite{Zhang:2021nnn}.
In terms of nonlinear evolutions, Ref.~\cite{Zhang:2021ybj} showed the energy transition between dilaton and the Maxwell field.
Besides, an effective repulsive force between them gives rise to a nontrivial scalar profile outside the black hole horizon.
Again, nonlinear dynamical evolution is shown crucial for gaining a deeper insight into scalarization and descalarization.
Therefore, it is worth generalizing the study of nonlinear dynamics in EMS theory to other physical systems containing scalar fields coupling with other physical sources.
This can help us further to understand the underlying physics in scalarization and descalarization and examine whether the current findings established for EMS theory are general. 
In this regard, the present study is largely motivated to explore the nonlinear dynamics in EdGB theory.
Furthermore, in the literature, most studies in the EdGB theory have been performed regarding either the linear perturbation or the decoupling limit~\cite{Blazquez-Salcedo:2016enn, Benkel:2016rlz, Benkel:2016kcq, Witek:2018dmd, Okounkova:2019zep}.
In~\cite{Blazquez-Salcedo:2016enn} the authors obtained the EdGB black hole and disclosed that they are linearly stable against both axial and polar perturbations.
By neglecting the backreaction of the scalar onto the spacetime, the scalar configuration in a Schwarzschild black hole was numerically investigated~\cite{Benkel:2016rlz}.
Ref.~\cite{Benkel:2016kcq} also studied dynamical evolution and the formation of scalar hair in a Schwarzschild spacetime and showed that the evolution eventually settles to static hairy solutions.
The stability of rotating black holes in EdGB theory was analyzed in Ref.~\cite{Okounkova:2019zep}, which indicated that the black holes are numerically stable up to the leading order.
Nonetheless, the linear perturbation and temporal evolution in the decoupling limit only probe the physics of the systems with a weak scalar field.
Moreover, it is unable to adequately address the backreaction of the scalar field on the gravity sector, such as how the scalar hair dynamically attaches to the bald black hole resulting in a hairy one, or how the hairy of the latter is deprived owing to relevant energy injection, among others.

To answer these questions, one needs to investigate the nonlinear far-from-equilibrium dynamics of scalarization and descalarization in the gravitational configurations.
In fact, some attempts have recently been made to study the nonlinear dynamics regarding theories containing scalar fields coupling with high curvature terms~\cite{Ripley:2019irj, Ripley:2019aqj, Ripley:2020vpk, Doneva:2021dqn,East:2020hgw,Corelli:2022phw}.
However, the emergence of ellipticity in particular spacetime regions was encountered~\cite{Ripley:2019irj, Ripley:2019aqj}.
In the system evolution in the EdGB theory, such an elliptic region typically appears at the exterior of the black hole horizon, indicating the break-down of the Cauchy problem.
In other words, it becomes rather delicate to establish a well-defined hyperbolic initial condition, and in practice, it causes the numerical code to crash when the system involves into such regions.
In Ref.~\cite{Antoniou:2021zoy}, a new action was suggested with scalar fields coupling with both GB term and Ricci scalar.
The analysis of linear perturbation~\cite{Andreou:2019ikc} to a Schwarzschild background showed that the onset of scalarization is determined entirely by the coupling with the GB term since the Ricci scalar vanishes.
However, compared with the GB term, the Ricci scalar term has a lower mass dimension and will contribute more to the effective mass.
Hence after the onset of scalarization, it will gradually dominate the nonlinear evolution that determines the properties of the final phase.
Moreover, Ref.~\cite{Antoniou:2022agj} investigated the effect of the Ricci coupling on the hyperbolicity of linear, radial perturbations, and showed such a theoretical framework is also advantageous in the sense that the above-mentioned elliptic region encountered in EdBG theory is reduced.
Therefore, in this work, we will employ this general coupling between scalar fields and GB term together with Ricci scalar.
We will investigate the reasons that ignite the scalarization and descalarization in the nonlinear physical process of scalar fields accretion onto isolated black holes.
In particular, a specific boundary condition will be adopted to avoid the nonphysical late-time decay of the scalar field, as encountered in some numerical studies of spontaneous scalarization.

\begin{figure}[tbp]
\centering 
\includegraphics[width=.6\textwidth]{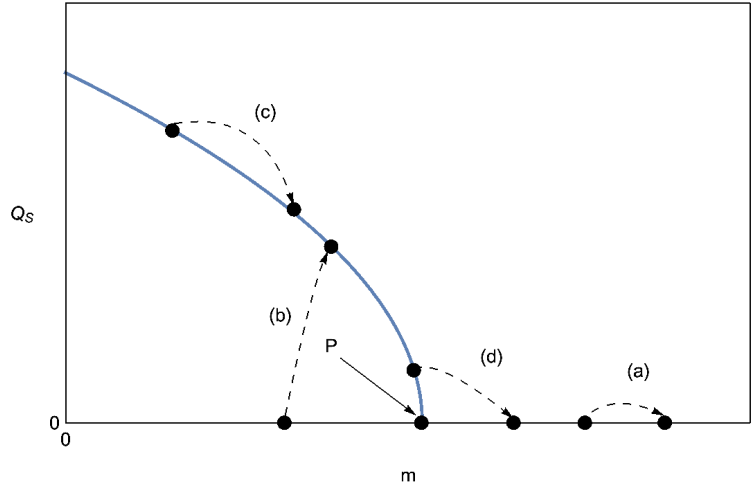}
\caption{\label{sketch} Schematic diagram for different far-from-equilibrium evolutions.
The vertical axis indicates the scalar charge, while the horizontal one denotes the Arnowitt-Deser-Misner (ADM) mass of the static HBH solutions.
The blue curve is the collection of static HBH solutions, specified below in Sec.~\ref{secHBH}.
The point $P$ on the horizontal axis indicates the value of the threshold mass.
An initial configuration with a mass above this threshold value will eventually evolve into a bald black hole, while a system with a mass below it will settle down to a hairy black hole.
As indicated in the figure, the dashed curves with arrows, labeled by (a) to (d), correspond to the four different scenarios explored in this study.
The arrows indicate the directions of temporal evolution. 
}
\end{figure}

Based on the above theoretical setup, we will explore the nonlinear dynamics of the system in an attempt to achieve a better understanding of the system's entire phase diagram.
In particular, it is rather intriguing to explore scenarios from one static stable to another static stable black hole spacetimes through the process of generation and removal of scalar hair.
As has been shown in EMS models~\cite{Zhang:2021nnn}, it is argued that the dynamic properties of a far-from-equilibrium system are not straightforward from the perspective of the linear stability analysis for static black hole solutions, in the sense that the final fate of the evolution could crucially depend on the strength of the initial perturbations.
We will start with bald black holes (BBHs) and investigate how they will evolve after the continuous absorption of scalar fields. 
We will reveal physical conditions, whether and how an initial BBH might remain bald and grow heavier, and how it can acquire scalar hair to become a hairy black hole (HBH). 
These relevant processes are described below by processes (a) and (b), respectively, in Fig.~\ref{sketch}. 
On the other hand, we will also study the evolution of HBHs regarding whether they may or may not keep the hair while swallowing scalar fields. 
As described by processes (c) and (d) below in Fig.~\ref{sketch}, an initial HBH can grow and become more massive, but instead, it may also lose its scalar hair and become bald. 
Why and how initial HBHs can have two drastically different fates worth careful investigations. 
We will illustrate in our study that nonlinear dynamics reflect the underlying physics that governs the different developments of initial BBHs and HBHs.

In the following sections, we give a detailed account of the numerical setup and simulations and elaborate further on the results regarding the far-from-equilibrium gravitational system.
The remainder of the paper is organized as follows. 
In Sec.~\ref{model}, we introduce the model and present the relevant equations of motion.
The properties of the static HBH solution are revisited in Sec.~\ref{secHBH}.
In Sec.~\ref{ch-initial-boundary}, the initial and boundary conditions are elaborated. 
The main results of the simulations are presented and discussed in Sec.~\ref{results}.
We address the evolution of the metric and the Misner-Sharp (MS) mass.
The last section is dedicated to further discussions and the concluding remarks.

\section{Action and equations of motion}\label{model}

We consider the black hole spacetime setup in terms of the following action
\begin{equation}\label{action}
S=\frac{1}{16\pi}\int d^{4}x\sqrt{-g}\left[R-\frac{1}{2}\partial_{\mu}\phi\partial^{\mu}\phi+W(\phi)\left(\frac{B}{2}R-A\mathcal{G}\right)-V(\phi)\right].
\end{equation}
Here $\mathcal{G}=R^2+R_{abcd}R^{abcd}-4R_{ab}R^{ab}$ is the Gauss-Bonnet invariant, 
where $R$ and $R_{ab}$ are the Ricci scalar and tensor, while $R_{abcd}$ is the Riemann tensor.
The parameter $B$ is dimensionless while $A$ has the dimensions of length squared, and $W(\phi)$ is a dimensionless coupling function.
Subsequently, the Einstein field equation reads
\begin{eqnarray}\label{eqs-motion}
G_{ab}&&=T_{ab},
\end{eqnarray}
where
\begin{eqnarray}
T_{ab}&&=\frac{1}{2}\nabla_a\phi\nabla_b\phi-\frac{1}{4}g_{ab}\nabla_c\phi\nabla^c\phi-\frac{1}{2}g_{ab}V(\phi)\nonumber\\
&&-A\delta^{cdef}_{ijkh}R^{kh}_{~~ef}\delta^{j}_{~(a}g_{b)d}\nabla^i\nabla_c W(\phi)-\frac{B}{2}(g_{ab}\nabla^2-\nabla_a\nabla_b+G_{ab})W(\phi),
\end{eqnarray}
where $\delta^{cdef}_{ijkh}$ is the generalized Kronecker delta tensor.
The equation of motion for the scalar field is given by 
\begin{equation}
\nabla_a\nabla^a \phi+W_{\phi}(\phi)(\frac{B}{2}R-A\mathcal{G})-V_{\phi}(\phi)=0,
\end{equation}
where the subscript `${\phi}$' denotes the derivative with respect to $\phi$.

By considering a spherically symmetric system, we write down the metric ansatz in Painleve-Gullstrand coordinates:
\begin{equation}\label{metric}
ds^{2}=-\alpha(t,r)^2dt^{2}+[dr+\alpha(t,r)\zeta(t,r)dt]^2+r^2(d\theta^2+sin^2\theta d\psi^2).
\end{equation}
For the Schwarzschild black hole $\alpha=1$ and $\zeta=\sqrt{2m/r}$.

To proceed further, one introduces the variables
\begin{equation}\label{auxF}
\Psi=\partial_r \phi,~~~~~~\Pi=\partial_t{\phi}/\alpha-\zeta \Psi ,
\end{equation}
so that the equations of motion can be rewritten as first order equations in $\phi, \Psi$ and  $\Pi$ as follows
\begin{eqnarray}
\partial_t \phi-\alpha (\Pi+\zeta \Psi)&&=0,~~~~~~~~~~~~~~~~~~~~~~\label{dynamiceq1}\\
E_{\Psi}\equiv\partial_t \Psi-\partial_r[\alpha(\Pi+\zeta \Psi)]&&=0,\label{dynamiceq2}\\
E_{\Pi}\equiv\partial_t \Pi-F(\Pi,\Psi,\phi,\alpha,\zeta)&&=0 ,\label{dynamiceq3}
\end{eqnarray}
where $F$ is an expression of $\Pi,\Psi,\phi,\alpha,\zeta$ and that possesses a rather tedious form. 

The remaining equations that only involve spatial derivatives serve as the Hamiltonian and momentum constraints.
They can be expressed in terms of the following partial differential equations in $\alpha$ and $\zeta$
\begin{eqnarray}
0=&&(1+A X+B Y)\zeta' +\left(\zeta+A\zeta X+B\frac{Y}{\zeta}\right)\frac{\alpha'}{\alpha}+A\frac{\zeta}{r}\left(\Psi^2 W_{\phi\phi}+W_{\phi} \Psi'\right)+\frac{\zeta}{2r} \nonumber\\
&&+B\left(\frac{W\zeta}{4r}-\frac{1}{2}P-\frac{\Psi W_{\phi} }{2\zeta}-\frac{\Psi^2r W_{\phi\phi} }{4\zeta}-\frac{r W_{\phi} \phi'}{4\zeta}\right)-\frac{r V}{4\zeta}-\frac{1}{8\zeta}(\Pi^2+\Psi^2),\label{eq-alpha}\\
0=&&\left[1+A\left(4\Psi+4P\zeta-2\Psi\zeta^2\right)\frac{W_{\phi} }{r}+B\left(\frac{W}{2}+\frac{\Psi r W_{\phi}}{4}\right)\right]\frac{\alpha'}{\alpha}+\frac{\Pi \Psi r}{4\zeta}\nonumber\\
&&-\frac{2A\zeta}{r}(\Pi\Psi W_{\phi\phi} +W_{\phi} P'+\Psi W_{\phi} \zeta')+\frac{Br}{4\zeta}(\Pi \Psi W_{\phi\phi} +W_{\phi} \Pi'+\Psi W_{\phi} \zeta'),\label{eq-zeta}
\end{eqnarray}
where the prime $'$ represents the derivative with respect to $r$.
The auxiliary functions are defined as 
\begin{eqnarray}\label{}
W_{\phi}=\frac{d}{d\phi} W(\phi),~~~W_{\phi\phi}=\frac{d^2}{d\phi^2}W(\phi),~~~X= \frac{4\Psi W_{\phi}}{r}+\frac{6\Pi\zeta W_{\phi}}{r},~~~Y=\frac{W}{2}-\frac{r\Pi W_{\phi}}{4\zeta}.\nonumber
\end{eqnarray}
In this paper, we consider the specific form $W(\phi)=-\phi^2/2$, and for simplicity the potential function is set to null $V=0$.
In the simulations, we take $A=1$ and $B=5$ without losing essential physical relevance.

\section{Static hairy black hole solutions}\label{secHBH}

This section briefly revisits the static HBH solutions to action Eq.~(\ref{action}).
Such a static solution was first derived in~\cite{Antoniou:2021zoy} in a diagonal metric.
Here, we numerically solve the model in Painleve-Gullstrand coordinates by manually turning off the time-dependence in the metric ansatz Eq.~(\ref{metric}).
The equations of motion can also be obtained by dropping the time-dependent terms in Eqs.~(\ref{dynamiceq3}), (\ref{eq-alpha}), and (\ref{eq-zeta}).
After some straightforward calculations, one obtains two coupled ordinary differential equations for $\phi$ and $\zeta$, as well as an algebraic equation for $\alpha$.
The boundary conditions are imposed as
\begin{eqnarray}
\alpha|_{r\rightarrow r_h}\simeq 1+a_1(r-r_h)+\dots, \nonumber
\end{eqnarray}
\begin{eqnarray}
\zeta|_{r\rightarrow r_h}\simeq 1+b_1(r-r_h)+\dots,\nonumber
\end{eqnarray}
and 
\begin{eqnarray}
\phi|_{r\rightarrow r_h}\simeq p_0+p_1 (r-r_h)+\dots \nonumber
\end{eqnarray}
on the black hole's horizon.
At infinity, the scalar field decays as 
\begin{eqnarray}
\phi\sim Q_s/r+\dots ,\nonumber
\end{eqnarray}
and the metric function approaches
\begin{eqnarray}
\zeta\sim \sqrt{2m/r}+\dots \nonumber
\end{eqnarray}
The coefficients $Q_s$ and $m$ are the spacetime's scalar charge and ADM mass, respectively.
It is noted that the scalar charge $Q_s$ here is not associated with any conservation law but is determined by the behavior of the scalar field at a significant distance.
For given $r_h$, we can obtain $\zeta$ and $\phi$ by the shooting method.
The metric function $\alpha$ is then obtained as an algebraic function of $\zeta$ and $\phi$.
After solving the equations, one can rescale $\alpha$ to have $\alpha|_{r\rightarrow \infty }\rightarrow 1$ at infinity by the gauge freedom.
We also note that the coordinates employed in Ref.~\cite{Antoniou:2021zoy} can be readily transformed into the static Painleve-Gullstrand coordinates, and the obtained results are manifestly consistent.

\begin{figure}[tbp]
\centering 
\includegraphics[width=.6\textwidth]{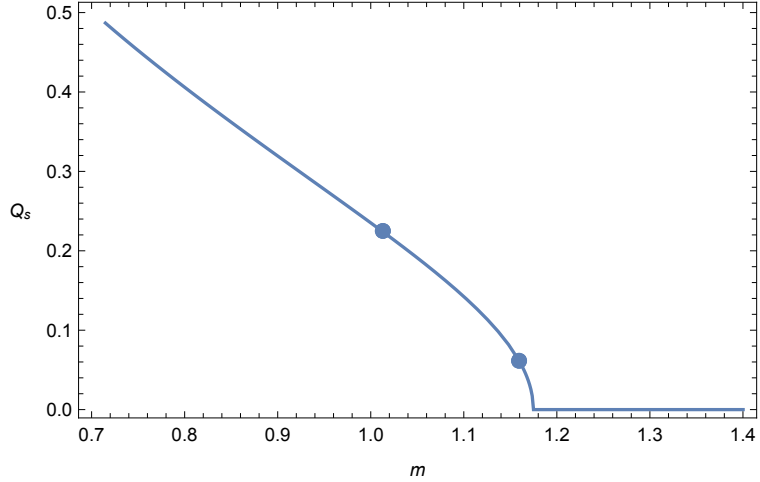}
\caption{\label{Qs-m} The relationship between the scalar charges $Q_s$ and ADM masses $m$ of the static HBH solutions numerically evaluated by the present model.
The profiles of two static HBH solutions indicated by filled circle and square on the curve are shown below in Fig.~\ref{s-s-initial}.
}
\end{figure}

\begin{figure}[tbp]
\centering 
\includegraphics[width=.31\textwidth]{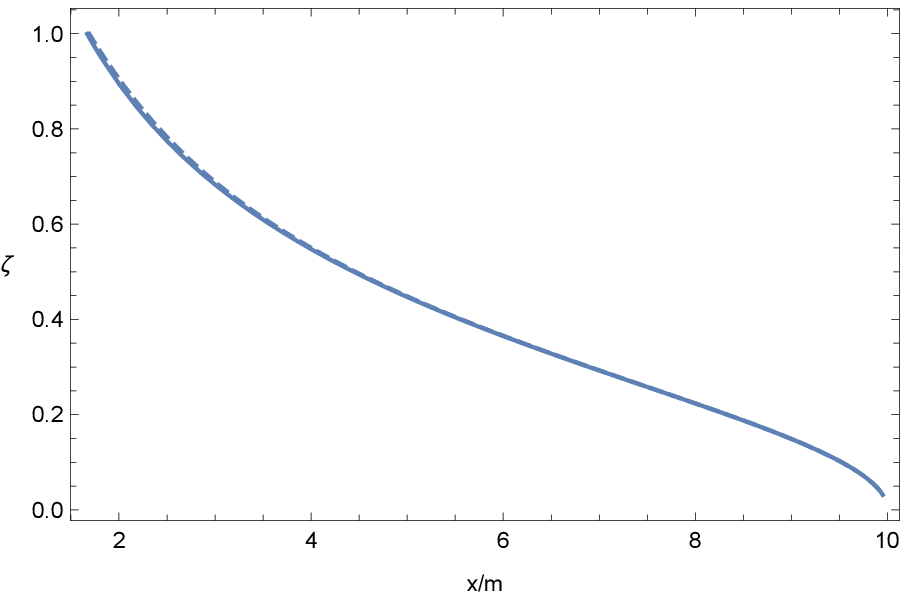}
\includegraphics[width=.31\textwidth]{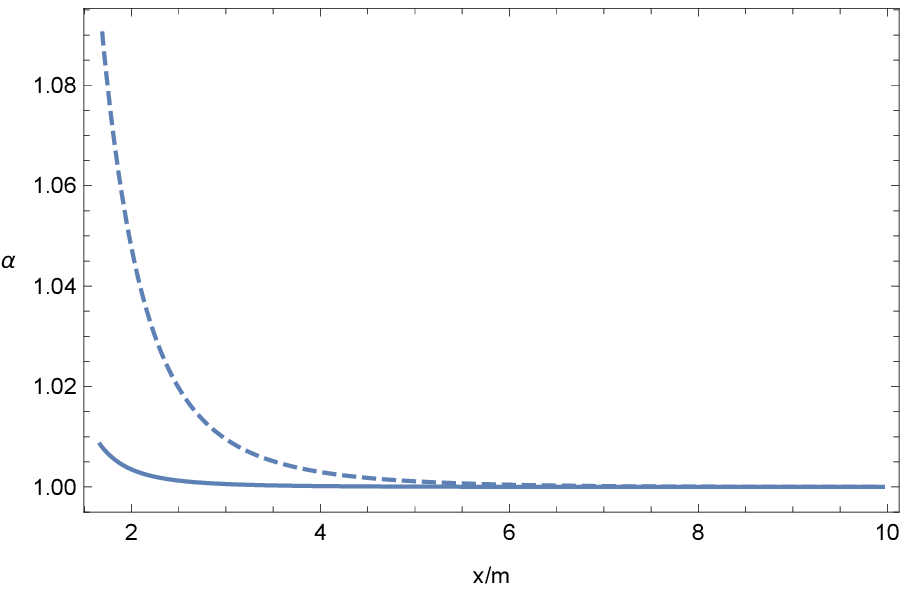}
\includegraphics[width=.31\textwidth]{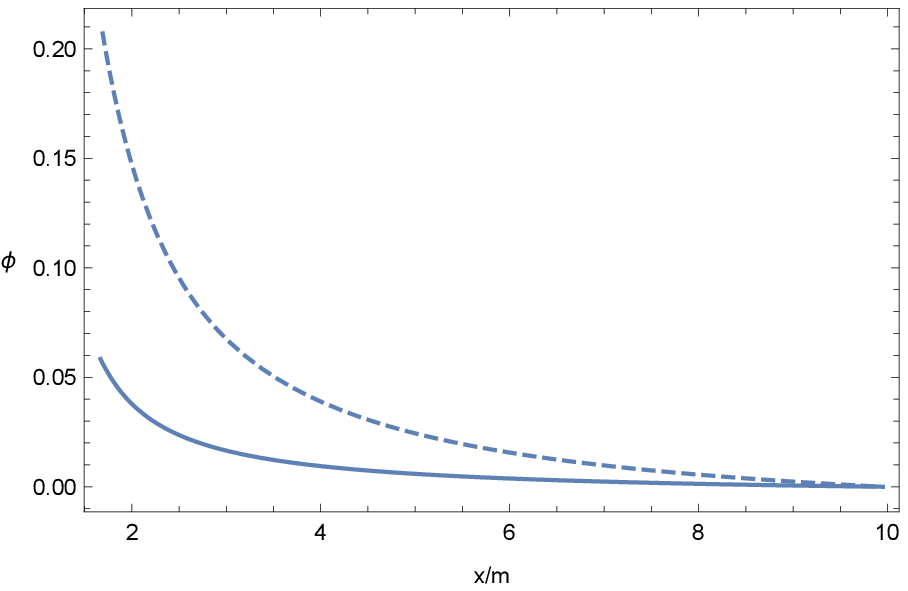}
\caption{\label{s-s-initial} 
The spacetime configurations of the static HBHs with $m=1.03424, Q_s=0.225$ (dashed curves) and $m=1.1594, Q_s=0.0615$ (solid curves), corresponding to the filled circle and square, respectively, in Fig.~\ref{Qs-m}.
The left, middle, and right panels show the metric functions, $\zeta(r)$ and $\alpha(r)$, as well as the scalar field $\phi(r)$ as functions of $x/m$.}
\end{figure}

In Fig.~\ref{Qs-m}, we show the relation between the scalar charge $Q_s$ and ADM mass $m$ of the static solutions where the scalar profile does not contain any node.
The relation indicated by the blue curve was presented schematically in Fig.~\ref{sketch}.
As the radial coordinate $r$ approaches infinity, the scalar field asymptotically approaches $\phi\sim Q_s/r+\dots$.
Numerically, the black hole's scalarization can be triggered when the mass is below a threshold value around $m_\mathrm{t}=1.1739457$, which is consistent with the result obtained in~\cite{Antoniou:2021zoy}.
For the black holes with $m < m_\mathrm{t}$, the HBH featured by a non-trivial scalar field distribution outside the horizon is the only stable solution.
On the other hand, the corresponding stable static black hole solution is a bald Schwarzschild one for heavier black holes whose mass is beyond this threshold.
The filled circle and square on the blue curve correspond to the HBH solutions with the parameters $m=1.03424, Q_s=0.225$ and $m=1.1594, Q_s=0.0615$, whose profiles of the metric functions and scalar fields are shown in Fig.~\ref{s-s-initial}.
In Fig.~\ref{s-s-initial}, we illustratively present the profiles for the metric functions $\zeta_b(r)$ and $\alpha_b(r)$, as well as the scalar field $\phi_b(r)$ of two static HBH solutions in solid (for the solution indicated by the filled square) and dashed (for that indicated by the filled circle) curves.

\section{Numerical Implementation}\label{ch-initial-boundary}

In this section, we proceed to investigate the dynamic aspects of the HBHs.
We discuss the characteristics of the differential equations, elaborate on the initial and boundary conditions, and comment on a few relevant details concerning the numerical methods.

\subsection{Characteristics}

Owing to the presence of the Gauss-Bonnet invariants in the action, such as the Horndeski and EdGB theories~\cite{Papallo:2017qvl,Papallo:2017ddx,Ripley:2019hxt, Ripley:2019irj, Ripley:2019aqj, Ripley:2020vpk}, the system of equations of motion that is initially hyperbolic could turn into an elliptic one at some particular spacetime region.
In such an elliptic region, the theory usually suffers from a ``Laplacian'' or ``gradient'' instability~\cite{Papallo:2017qvl, Papallo:2017ddx}, which demonstrates that the theory Eq.~(\ref{action}) does not admit a well-defined hyperbolic initial condition.
In practice, to prevent the code from breaking down, we will cast out the elliptic region from the domain of our numerical calculations.
To this end, we proceed to study the characteristics of the system of equations for the relevant degrees of freedom in order to monitor the entrance of the elliptic region.
For a detailed pedagogical review on the topic, the reader is referred to~\cite{courant1962methods, kreiss1989initial, gustafsson1995time}.

We first algebraically solve the constraint equations (\ref{eq-alpha}) and (\ref{eq-zeta}) for $\partial_r \alpha$ and $\partial_r \zeta$, and then substitute the obtained expressions into Eqs.~(\ref{dynamiceq2}) and~(\ref{dynamiceq3}).
Subsequently, the two radial characteristic speeds of the scalar field $\phi$ for, respectively, the ingoing and outgoing modes, $c_{\pm}=\mp \xi_{t}/\xi_{r}$, are obtained by solving the characteristic equation
\begin{eqnarray}\label{char}
\mathrm{det}\left[\left(\begin{array}{ccc} \delta E_{\Pi}/\delta (\partial_{\mu}\Pi) &  \delta E_{\Pi}/\delta (\partial_{\mu}\Psi)  &  \\ \delta E_{\Psi}/\delta (\partial_{\mu}\Pi)  &  \delta E_{\Psi}/\delta (\partial_{\mu}\Psi)  \\ \end{array}\right)\xi_{\mu}\right]=0.
\end{eqnarray}
It is noted that Eq.~(\ref{char}) is of quadratic form $a x^2+b x+ c=0$ for the radial speeds, where the coefficients $a, b$ and $c$ are governed by the metric functions $\alpha$, $\zeta$ and scalar field $\phi$.
The sign of the discriminant $D=b^2-4ac$ determines the character of the system of partial differential equations Eqs.~(\ref{dynamiceq2}-\ref{dynamiceq3}).
The spacetime region for which $D>0$ corresponds to the domain where the characteristic speeds are real and therefore, Eqs.~(\ref{dynamiceq2},\ref{dynamiceq3}) are hyperbolic.
On the other hand, in the region where with $D<0$, the characteristic speeds become complex (with a non-vanishing imaginary component), and subsequently, Eqs.~(\ref{dynamiceq2},\ref{dynamiceq3}) are elliptic.
These two regions are separated by the boundary defined by $D=0$, where the equations of motion are parabolic.

The black hole's apparent horizon can be identified dynamically by allocating the marginally trapped surface formed by null outgoing characteristics~\cite{Thornburg:2006zb}.
In Painleve-Gullstrand coordinates, the location of such surface is determined by $\zeta(t,r_h)=1$ with $r_h$ being the horizon radius.
For the spherically spacetime at hand, we use the MS mass to define the quasi-local mass by 
\begin{eqnarray}\label{ms-mass}
M_\mathrm{MS}(t,r)=r\zeta(t,r)^2/2,
\end{eqnarray}
where $r$ is the areal radius.
It gives the total mass of the spacetime at the limit $r\rightarrow \infty$.
Moreover, one can relate the MS mass to the charge associated with the Kodama current~\cite{Abreu:2010ru}.
The MS mass corresponds to an integral of the effective matter energy density using Eq.~\eqref{eqs-motion}.
It is noted that the effective stress tensor $T_{\mu \nu}$ does not always satisfy the usual energy condition as the MS mass here is not necessarily a monotonically increasing function of radial radius $r$.
We will elaborate further on the non-monotonicity of $M_\mathrm{MS}$ below.

\subsection{Initial and boundary conditions}

This section discusses the initial and boundary conditions of the numerical calculations regarding the dynamic evolution of the system.
The physical picture here is to introduce an energy injection into an initially static black hole spacetime.
The latter can be either a bald Schwarzschild black hole or a scalarized spherically symmetric hairy one, hereafter denoted by BBH and HBH, respectively, in the remainder of the paper.
We also denote the background metric of the initial spacetime by $\zeta_b(r)$ and $\alpha_b(r)$, and the initial scalar profile by $\phi_b(r)$.
For an initially BBH, we have $\alpha_b(r)=1$ and $\zeta_b(r)=\sqrt{2m/r}$ with the mass $m$, while $\phi_b(r)$ vanishes. 
For an initial HBH, the initial static solutions of the metric functions are furnished numerically.

In addition to the background scalar field $\phi_b$, one introduces an energy injection in terms of the initial perturbations denoted by $\phi_p$, that is located within the range $r_l<r<r_u$:
\begin{eqnarray}\label{initial-scalar}
\phi_p(r)=\left\{
\begin{array}{rcl}
&&a_m(r-r_l)^2(r_u-r)^2\exp[-\frac{r}{r-r_l}-\frac{1}{r_u-r}] ~~~r_l<r<r_u,\\
&&0~~~~~~~~~~~~~~~~~~~~~~~~~~~~~~~~~~~~~~~~~~~~~~~~~~~~~~~~~~ r\le r_l,~r\ge r_u ,
\end{array}
\right.
\end{eqnarray}
where the parameter $a_m$ plays the role as the amplitude of the scalar perturbation, and we hereafter take $r_l=24$ and $r_u=32$.
The initial pulse of the scalar field is spherical symmetric distribution without any node. 
One further assumes the initial perturbations are static. 
Using Eq.~\eqref{auxF}, the initial auxiliary variables are given by 
\begin{eqnarray}\
\Psi(t=0,r)&=&\partial_r\phi_b(r)+\partial_r\phi_p(r),\nonumber\\
\Pi(t=0,r)&=&-\zeta_b(r)\partial_r\phi_b(r).
\end{eqnarray}

We employ a stereographic projection to compactify the $r$ coordinate as $x=\frac{r}{1+r/L}$~\cite{Ripley:2020vpk}. 
In the radial direction, on each time slice, the physically pertinent range $r\in [r_{e},+\infty)$ corresponds to the interval $x\in[x_{e}, L]$.
In our calculations, we assume the value $L=10 m$.
Also, we note that the excision position $r_e = r_e(t)$ is time-dependent, which marks the edge below which the system of equations is elliptic, and subsequently, the corresponding space domain is excised.
Since the region inside the black hole horizon is irrelevant for the present study, on the initial time slice, we set the excision coordinate $r_e$ to be the initial black hole horizon radius.
By specifying the energy injection $\phi(t=0,r(x))$, $\Psi(t=0,r(x))$, and $\Pi(t=0,r(x))$, we solve the constraints Eqs.~(\ref{eq-alpha}-\ref{eq-zeta}) in the domain $x\in[x_{e},L]$ using Newton's method in conjunction with the boundary conditions.
Due to the gauge freedom $\alpha\rightarrow l \alpha, ~t\rightarrow t/l$ in the metric Eq.~(\ref{metric}), the latter are taken to be $\alpha(t=0,r(x_e))=1$ and $\zeta(t=0,r(x_e))=\zeta_b(r(x_e))$, defined at $x_e$ and $t=0$.
Subsequently, one obtains the metric functions $\alpha(t=0,r(x))$ and $\zeta(t=0,r(x))$ for arbitrary $x\in[x_{e}(0),L]$ on the first time slice.
By substituting $\zeta(t=0,r(x))$ back into Eq.~(\ref{ms-mass}), one obtains the MS mass of the entire spacetime at the initial time $M_{\infty}=M_\mathrm{MS}(t=0,+\infty)$.
We note that the initial excision point and the horizon is merely a choice, that does not leads to any physical implication.

For the subsequent time slices, we solve for the metric functions $\alpha$ and $\zeta$ by imposing the boundary conditions at the spatial infinity.
Again, we take $\alpha(r\rightarrow\infty)=1$ by exploiting the gauge freedom.
Also, to improve the stability of the solution $\zeta$ at large $r$ (or $x$), one introduces a new variable $\eta=\zeta r^{1/2}$.
The boundary condition for $\eta$ at infinity is then recognized to be $\sqrt{2M_{\infty}}$ which is a conserved quantity during the temporal evolution.

To evolve the variables $\phi$, $\Pi$ and $\Psi$ in time using Eqs.~(\ref{dynamiceq2}-\ref{dynamiceq3}), we use a fourth-order finite difference method to evaluate spatial derivatives and a fourth-order Runge-Kutta method in the time direction. 
Once $\alpha$, $\zeta$ $\phi$, $\Pi$, and $\Psi$ are obtained, one utilizes Eq.~(\ref{char}) to monitor the radial characteristics. 
If the elliptic region forms, one casts it out from the domain of interest and redefines the range $[x_e, L]$ for the next time slice.
In practice, the numerical calculations turn out to be rather time-consuming for the initial scalar pulse to reach the spacial infinity.
The Courant–Friedrichs–Lewy condition is set to $0.25$ in our simulations.
It is also worth mentioning that the scalar field falls off at infinity according to the form $\phi \sim Q_{s}/r$, and therefore we take the value of $r\phi(r)$ at the large $r$ as the scalar charge.

\section{Numerical Results}\label{results}

In this section, we present the numerical results on the dynamics of the gravitational systems, such as the evolution of the metric functions, scalar charge, and the MS mass, and discuss their physical relevance.
As shown above, in Fig.~\ref{sketch} of Sec.~\ref{intro}, we will consider four different scenarios, elaborated in the following subsections.
For an initially BBH, the metric function are taken as $\alpha_b(r)=1$ and $\zeta_b(r)=\sqrt{2m/r}$.
Depending on the mass of the initial black hole, when compared to the threshold value, the initial scalar perturbations might trigger the gravitational system to evolve into different final states.
For an initial HBH, we also consider two different values for the initial black hole mass.
As will be demonstrated below, using an appropriate amount of energy injection, the system might evolve into either an HBH with less scalar hair or a BBH with a mass slightly more significant than the threshold value $m_\mathrm{t}$.
The latter case corresponds to a process where the hair of an HBH is deprived, and the system eventually settles down to a BBH.
For the parameters concerned in our work, the hyperbolicity of the equations for the physically relevant region remains intact during the process of evolution, and the excision, therefore, does not lead to any physical implication.

\subsection{Route (a): BBH to BBH}

We first present the results on black hole evolution as indicated by route (a) introduced in Sec.~\ref{intro}.
Without loss of generality, we take the mass of the initially static black hole to be $m=1.3$ beyond the threshold value.
The amplitude of the scalar perturbations is chosen to be $a_m=3\times 10^{-5}$.
As a result, the initial scalar perturbations contribute a fraction $\delta m_{\phi}/m\simeq 10^{-3}$ of the total mass of the spacetime.
The calculations are carried out by using $n=2^{11}+1$ spatial grids.

\begin{figure}[tbp]
\centering 
\includegraphics[width=.38\textwidth]{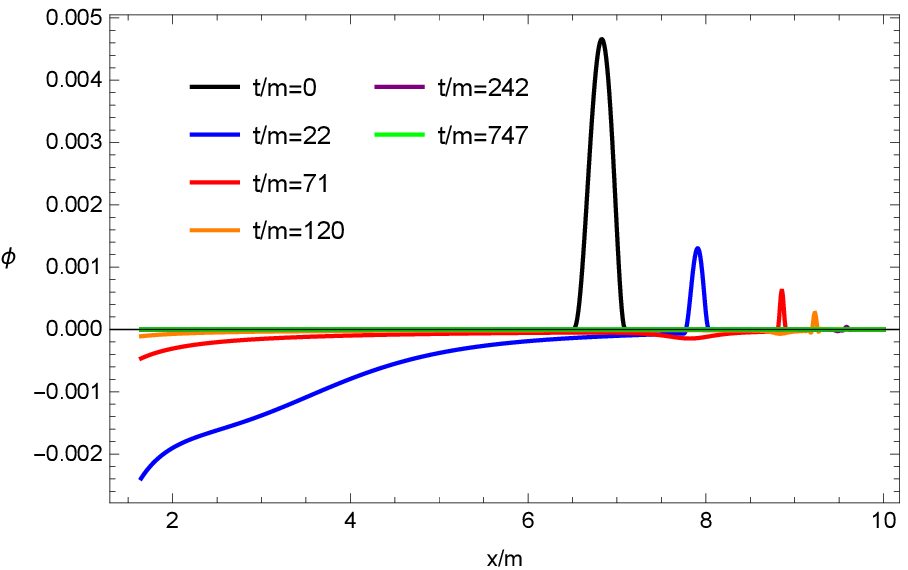}
\includegraphics[width=.383\textwidth]{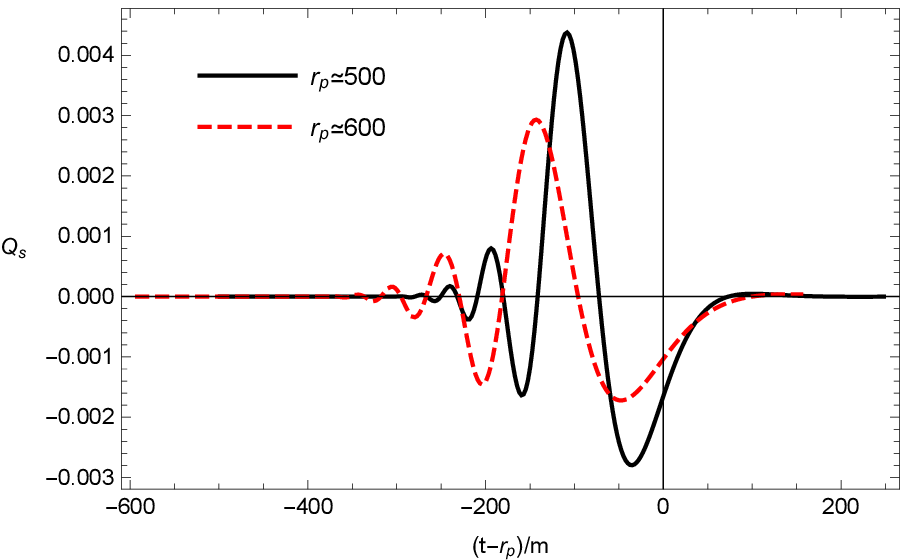}
\caption{\label{phi-decay} The temporal evolutions of the scalar field for an initial BBH with a mass beyond the threshold value.
Left panel: spatial distribution of the scalar field $\phi$ at different time slices.
Right panel: the scalar charge $Q_s$ as a function of $t-r_p$ at large $r_p$. }
\end{figure}

The numerical results are shown in Fig.~\ref{phi-decay}.
The left panel gives the spatial profiles of the scalar field at various time instants during the evolution, and the right panel shows the scalar charge as a function of time, evaluated at given spatial coordinates $r_p=500$ and $r_p=600$.
At the initial stage of the dynamical process, part of the scalar field propagates towards the black hole horizon as the magnitude of the scalar field temporarily increases in the vicinity of the horizon.
Subsequently, as the scalar field interacts with the black hole, it is absorbed by the horizon.
As a result, the strength of the field decreases in time near the horizon and eventually vanishes.
On the other hand, a small fraction of the scalar perturbation moves outward. 
Meanwhile, the magnitude of the initial pulse is suppressed in time until it entirely disperses to infinity.
From the left panel of Fig.~\ref{phi-decay}, we see that at $t/m=242$ scalar field mostly vanishes over the entire spatial range, which is almost identical to the profile at the later time $t/m=747$.
The above results are obtained for calculations using different magnitudes of the perturbations and initial masses beyond the threshold.
Therefore, we conclude that the initial scalar field perturbations vanish for the first scenario by either being absorbed by the horizon or dispersing to infinity.
The resultant spacetime settles down to a Schwarzschild BBH.
We noted that this is consistent with the results shown in the right panel of Fig.~\ref{phi-decay}.
The given coordinates $r_p=500$ and $r_p=600$ can be numerically consider as ``infinity''.
The displacement $r_p$ in time ensures that the peak of the curve would be placed near the origin.
One observes that the scalar charge $Q_s$, as a function of time, asymptotically vanishes at both infinities.
As the perturbations are initially static, it gives rise to a mostly vanishing initial scalar charge.
During the evolution, the scalar charge oscillates.
As the scalar field dissipates, the scalar charge asymptotically vanishes, indicating that the black hole's final state is free of any scalar hair.

\subsection{Route (b): BBH to HBH}

We proceed to discuss the scenario regarding the dynamic process of scalarization as the route (b) indicated in Sec.~\ref{intro}.
In order to illustrate the results, we take the mass of the initial BBH to be $m=1$, below the threshold value.
Again, the amplitude of the scalar perturbations is chosen to be $a_m=3\times 10^{-5}$.
Also, the calculations are carried out by using $n=2^{11}+1$ spatial grids.

\begin{figure}[tbp]
\centering 
\includegraphics[width=.38\textwidth]{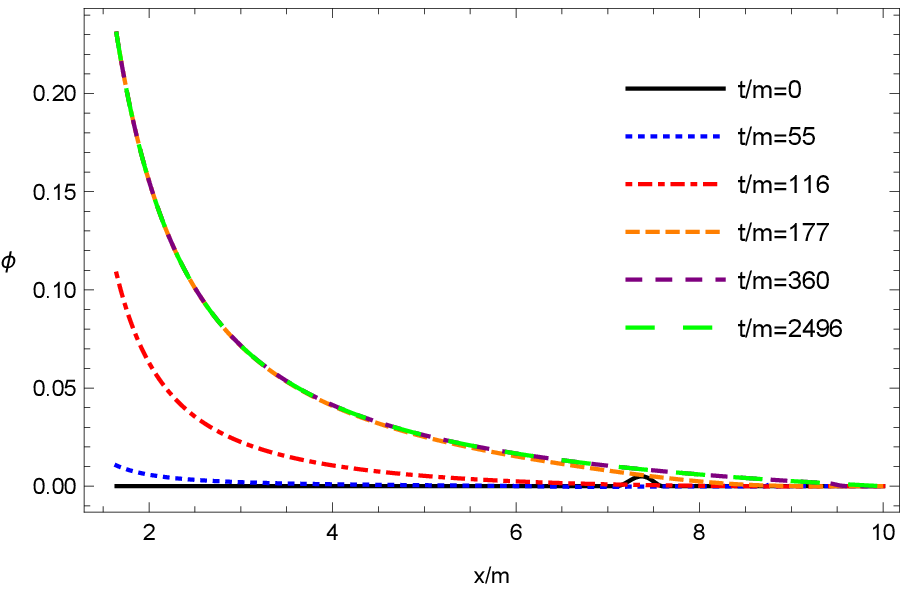}
\includegraphics[width=.383\textwidth]{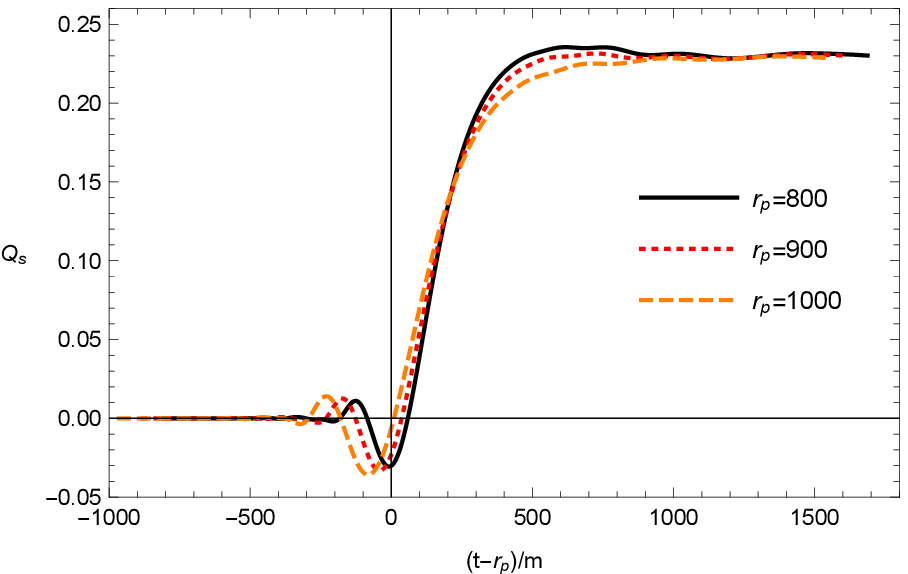}
\caption{\label{phi-grow} 
The temporal evolutions of the scalar field for an initial BBH with a mass below the threshold value.
Left panel: spatial distribution of the scalar field $\phi$ at different time slices.
Right panel: the scalar charge $Q_s$ as a function of $t-r_p$ at large $r_p$.  }
\end{figure}

\begin{figure}[tbp]
\centering 
\includegraphics[width=.32\textwidth]{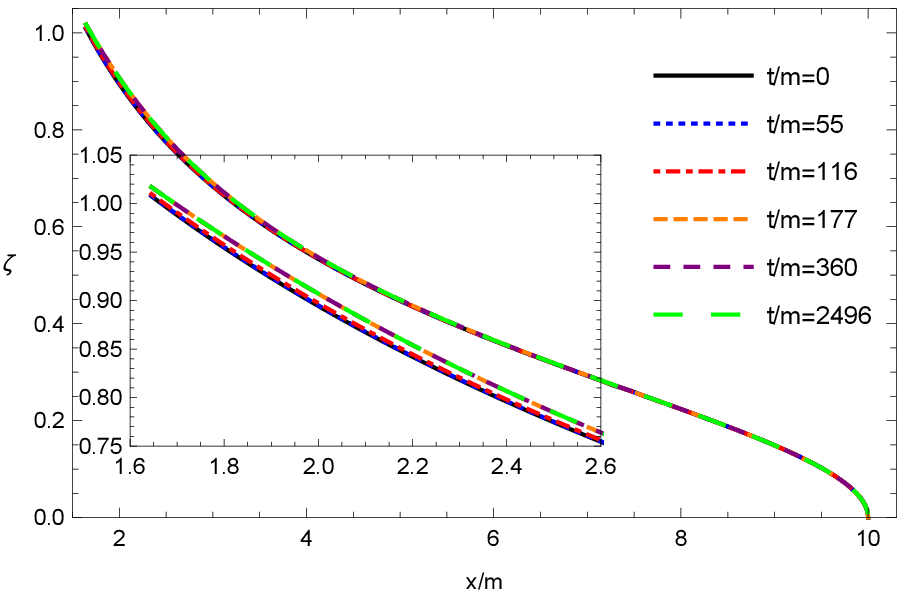}
\includegraphics[width=.335\textwidth]{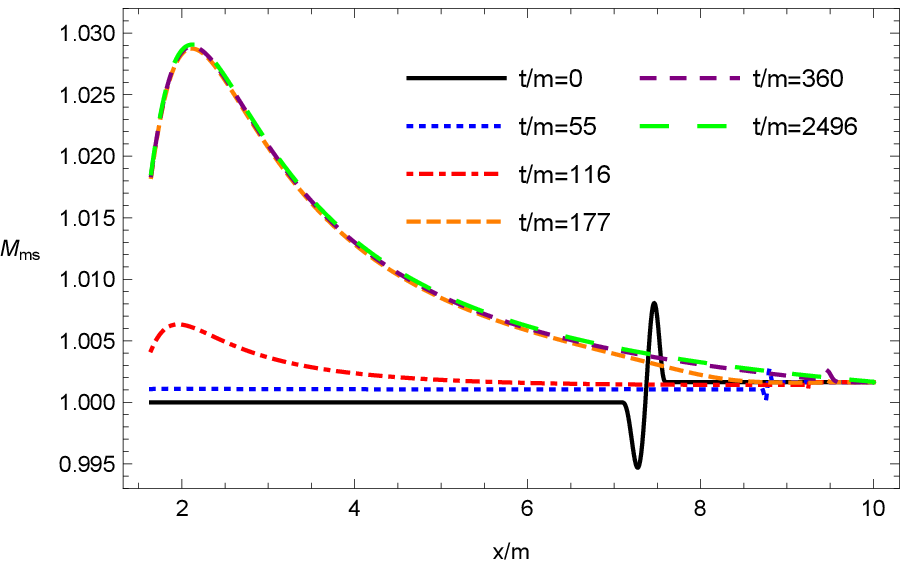}
\includegraphics[width=.33\textwidth]{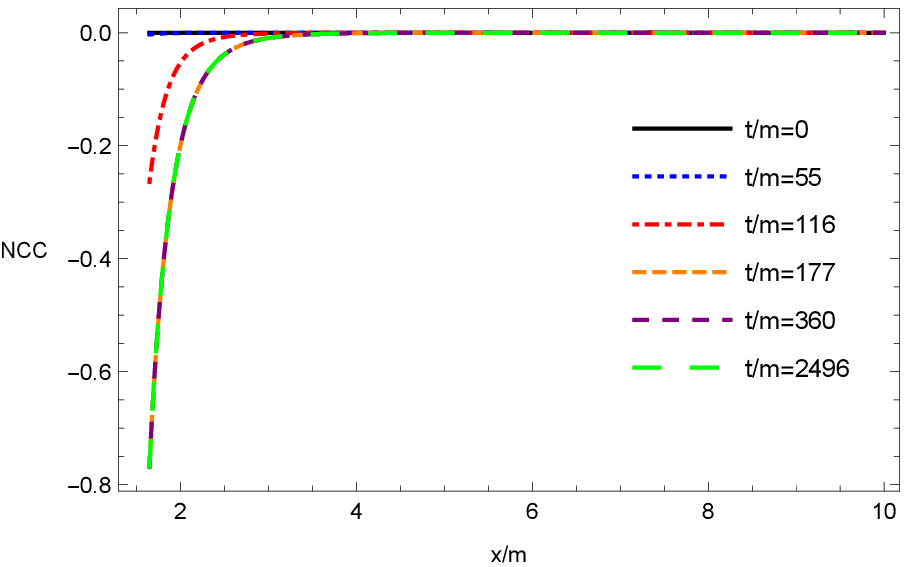}
\caption{\label{zeta-mass-grow} 
The calculated evolutions of the metric function $\zeta$ (left), MS mass $M_\mathrm{MS}$ (middle) and $f = R_{\mu\nu} l^\mu l^\nu$ (right) as functions of the rescaled radial coordinate $x/m$ for various time slices.}
\end{figure}

\begin{figure}[tbp]
\centering 
\includegraphics[width=.35\textwidth]{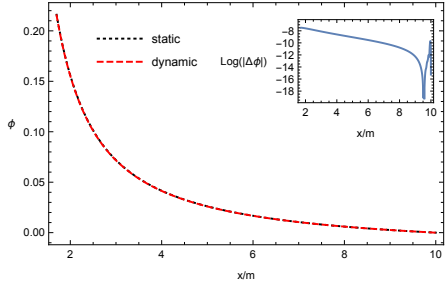}
\caption{\label{phi-diff} The spatial profile of the scalar field $\phi$, as a function of $x/m$, for the static HBH solution, compared against that as the final state resulting from the dynamic scalarization process.
The inset gives the difference $\Delta\phi$.
}
\end{figure}

The numerical results are shown in Figs.~\ref{phi-grow}-\ref{phi-diff}.
In Fig.~\ref{phi-grow}, the left panel gives the spatial profiles of scalar field at various time instants during the evolution, and the right panel show the scalar charge as a function of time, evaluated at given spatial coordinates $r_p=800$, $r_p=900$, and $r_p=1000$.
Triggered by the initial perturbations, the strength of the scalar field grows in time in the vicinity of the horizon.
The scalar field saturates and converges to a well-defined distribution as the time increases. 
Numerically, the solution becomes time independent before $t/m=177$, as the curves at the time instants  $t/m=360$ and $2496$ overlap entirely with that at $t/m=177$.
The dynamic process is recognized as spontaneous scalarization, and the resultant scalarized black hole solution is a static HBH.
On the other hand, one may also investigate the evolution of the scalar charge.
This is evaluated and shown in the right panel of Fig.~\ref{phi-grow} at three different radial coordinates $r_p\approx 800, 900$, and $1000$.
It is observed that the scalar charge oscillates and then increases in time until it eventually converges to a given value $Q_{s} \approx 0.23$.
The latter is consistent among the results evaluated at different radial coordinates.

From the left plot of Fig.~\ref{zeta-mass-grow}, one observes that the spacetime metric remains unchanged mainly during the scalarization.
This is partly attributed to the fact that the energy associated with the initial scalar perturbation counts only a tiny fraction, approximately the order of $10^{-3}$, of the mass of the initial BBH.
From the inset of the left panel, we see that $\zeta$ slightly increases over the course of the scalarization.
This implies that the radius of the apparent horizon, which satisfies $\zeta(t,r)=1$, also increases slightly, in agreement with the black hole area theorem.
The middle panel shows the temporal evolution of the MS mass $M_\mathrm{MS}(t,x)$.
The MS mass evaluated at the horizon, $M_\mathrm{MS}(t,x_h)$, indicates the mass of the black hole.
This implies that the black hole mass increases and saturates before the instant $t/m=177$, consistent with the behavior of the black hole horizon.
Also, the derivative of $M_\mathrm{MS}(t,r)$ with respect to the radial coordinate $r$ could be associated with the energy density of the scalar field.
For a given time slice, $M_\mathrm{MS}(t,r)$ is not a monotonical function. 
Therefore, it implies that the energy density of the scalar field outside the black hole horizon is not positive everywhere.

The null convergence condition (NCC) is the geometric interpretation of the null energy condition that requires $R_{\mu\nu} l^{\mu}l^{\nu}\geqslant 0$ for an arbitrary null vector $l^{\mu}$.
It plays an essential role in the laws of black hole mechanics and dynamical horiozn~\cite{Bardeen:1973gs, Hayward:1993wb, Ashtekar:2004cn}, and the topological censorship theorems~\cite{Friedman:1993ty, Galloway:1999bp}.
For the in-going null vector $l^{a}=(1,-\alpha(t,r)(1+\zeta(t,r)),0,0)$, the NCC implies $f=R_{\mu\nu} l^{\mu}l^{\nu}=2\alpha[\partial_t \zeta+(\zeta+1)^2\partial_r\alpha]/r \ge 0$.
The right panel of Fig.~\ref{zeta-mass-grow} shows the the scalar $f=R_{\mu\nu} l^{\mu}l^{\nu}$ at different time slices.
We note that the region where NCC is violated roughly coincides with where the energy density becomes negative.
Mathematically, by definition, the NCC or the $tt$ component of the effective energy-momentum tensor (c.f. Eq.~(26) of~\cite{Ripley:2019irj}, also see discussions in Refs.~\cite{Kanti:2011jz, Chakrabarti:2017apq, Ripley:2019aqj, Antoniou:2019awm}) receive a contribution associated with the Gauss-Bonnet scalar that might become negative, which plays an essential role during the spontaneous scalarization process.
The above results are general for BBH's initial state, where the initial black hole mass is below the threshold value.

In Fig.~\ref{phi-diff}, we compare the obtained spatial profiles of the scalar field between the static HBH solution and the asymptotic solution as the final state of the dynamic scalarization process.
The two profiles are in satisfactory agreement.
In the inset we show the difference of $\Delta\phi=|\phi_{static}-\phi_{dynamical}|$, which is less than $e^{-7}$. 
This coincidence shows that the far-from-equilibrium evolution indeed converges to the static scalarized black hole as the final stage of the evolution.
On the other side, it ascertains the precision of our numerical implementation.

\subsection{Route (c): HBH to HBH}\label{c}

The third scenario deals with an initial HBH whose mass is (sufficiently) smaller than threshold value.
Without loss of generality, the following results are obtained by taking value $m=1.03424$.
Again, one perturbs such a spacetime configuration by introducing scalar perturbations governed by Eq.~(\ref{initial-scalar}) with an amplitude $a_m=3\times 10^{-5}$.
In particular, we utilize the static solution presented by the dashed curves in Fig.~\ref{s-s-initial} as the initial configuration of HBH.
The strength of the scalar field at the horizon reads $\phi_h=0.18$, and the scalar charge is found to be $Q_s=0.2045$.

\begin{figure}[tbp]
\centering 
\includegraphics[width=.32\textwidth]{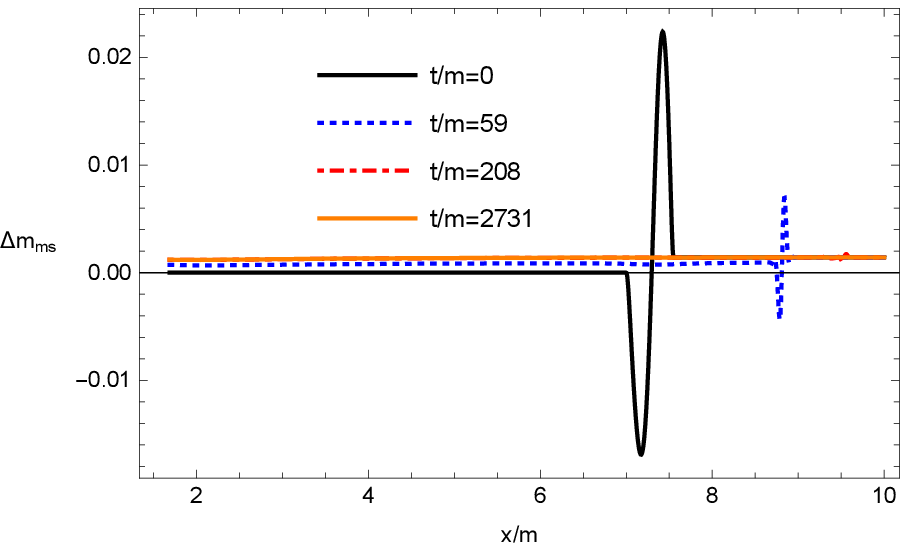}
\includegraphics[width=.32\textwidth]{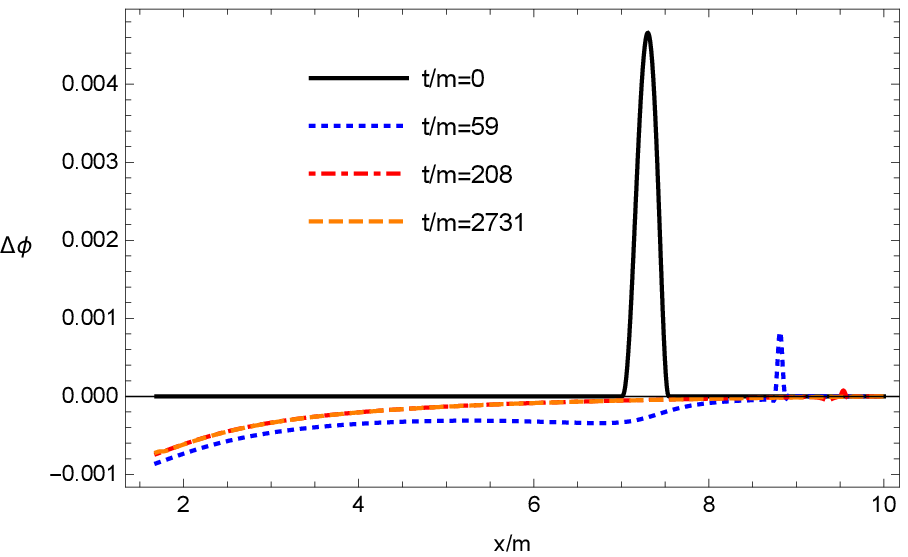}
\includegraphics[width=.32\textwidth]{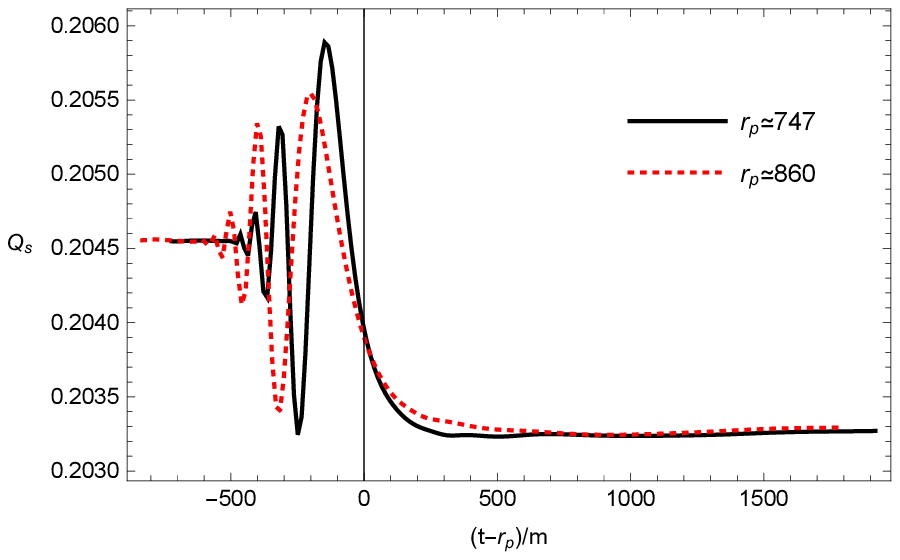}
\caption{\label{s-s-phi} 
Left: the deviation of the MS mass as functions of $x/m$, evaluated at different time instants.
Middle: the deviation of the scalar field as functions of $x/m$, evaluated at different time instants.
The deviations are evaluated with respect to their values of the initially static HBH.
Right: the scalar charge as a function of $(t-r_p)/m$, evaluated at given radial coordinates $r_p=747$ and $860$.
}
\end{figure}

We note that the MS mass of the entire spacetime can be viewed as a summation of two parts, namely, the mass of the black hole and that of the scalar field.
In this regard, during the dynamic evolution triggered by the perturbations of the scalar field, the total MS mass is modified due to two factors, namely, the variation of the MS mass of the black hole and that of the scalar field.
Therefore, to present our results more transparently, we will show the deviation of the MS masses associated with the black hole and scalar field from their respective values of the initially static HBH solution.
To be specific, we compute $\Delta M_\mathrm{MS}=M_\mathrm{MS}(t,x)-M^\mathrm{MS}_b(x)$, $\Delta \phi=\phi(t,x)-\phi_b(x)$, where $M_b^\mathrm{MS}$ is the MS mass of the initially static HBH spacetime.
These results are shown in Fig.~\ref{s-s-phi}.

At the beginning of the evolution, $t/m=0$, $\Delta M_\mathrm{MS}$ vanishes near the horizon, as expected. 
In other words, in this region, the geometry is determined by the black hole, which is consistent with the vanishing $\Delta \phi$ in the vicinity of the horizon shown in the middle panel.
It is also in accordance with the observation that $\Delta M_\mathrm{MS}$ oscillates in the region where the initial pulse of the scalar perturbations is injected.
As the radial coordinate goes to infinity, $\Delta M_\mathrm{MS}$ approaches a constant, which corresponds to the total MS mass of the entire initial spacetime in the presence of the scalar perturbations.
Numerically, the difference $\Delta M_\mathrm{MS}(t=0,\infty)$ evaluated at infinity is a positive but small value.
This indicates that the scalar perturbations give a positive overall correction to the mass of the entire spacetime, although the mass density, which is proportional to the derivative of MS mass in $r$, could become negative in specific regions.

As time proceeds, the MS mass evaluated at the horizon increases slightly.
This can be attributed to the absorption of a small fraction of the scalar field by the black hole horizon. 
On the other hand, the initial perturbations in the MS mass dissipate while propagating towards the spatial infinity.
The above observation regarding the propagation of the initial pulse is confirmed by the evolutions of the scalar profiles $\Delta \phi$ as presented in the middle panel of Fig.~\ref{s-s-phi}.
At an early instant, approximately $t/m=59$, the scalar field near the horizon decreases and attains the most minor (positive) value (while the deviation is negative).
Then, it slightly bounces back and saturates before $t/m=208$.
For the final state HBH solution, the condensation of the scalar field becomes less significant when compared with the initial HBH metric.
This is, again, in accordance with the asymptotic value of the scalar charge $Q_s$ evaluated in the right panel.
After some initial oscillations, the scalar charge converges to a smaller value $Q_s=0.2033$.
The calculations carried out by choosing different values of $r_p$ are in reasonable agreement.


\subsection{Route (d): HBH to BBH}

This subsection explores the scenario where a stable HBH loses its scalar hair due to scalar perturbations. 
To this end, we consider an initial HBH with the mass $m=1.17379$, which is slightly below the threshold $m_\mathrm{t}$.
The scalar pulse is then introduced according to the form given by Eq.~(\ref{initial-scalar}), where the amplitude is taken to be $a_m=9\times10^{-5}$.
During the process, the scalar hair of the HBH is deprived.
Meanwhile, the mass of the resultant spacetime increases and converges to $m=1.18661$, slightly above the threshold.
Therefore, the resulting BBH is stable, consistent with the findings in~\cite{Antoniou:2021zoy}.
One observes that the scalar pulse contributes about one percent of the mass of the entire spacetime.

The initial configuration of metric functions $\zeta_b(r)$ and $\alpha_b(r)$, and the scalar field $\phi_b(r)$ of the static HBH (not shown) are mainly similar to those presented above in Fig.~\ref{s-s-initial}.
It is noted that in Painleve-Gullstrand coordinates, for a BBH metric, one has $\alpha=1$.
From the scalar field distribution shown in the right panel, one finds the scalar charge to be $Q_s = 0.0064$.


\begin{figure}[tbp]
\centering 
\includegraphics[width=.32\textwidth]{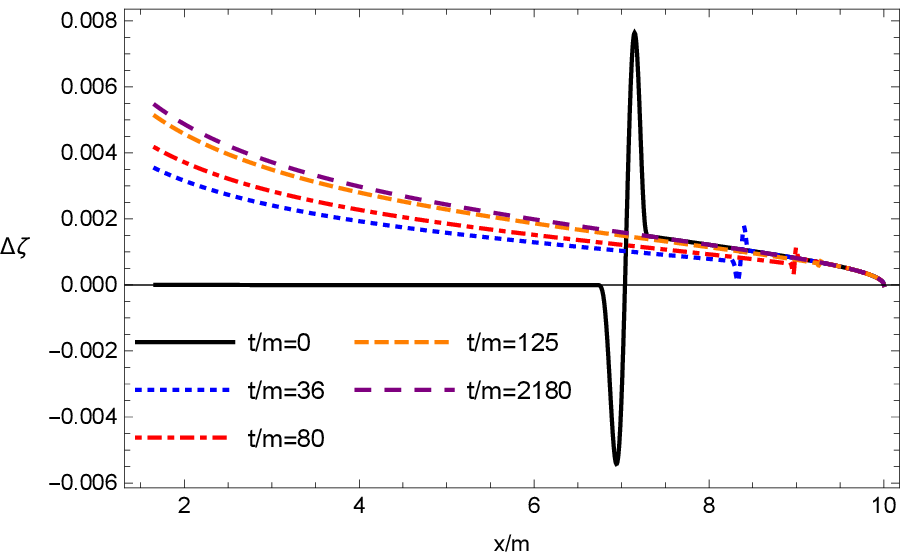}
\includegraphics[width=.335\textwidth]{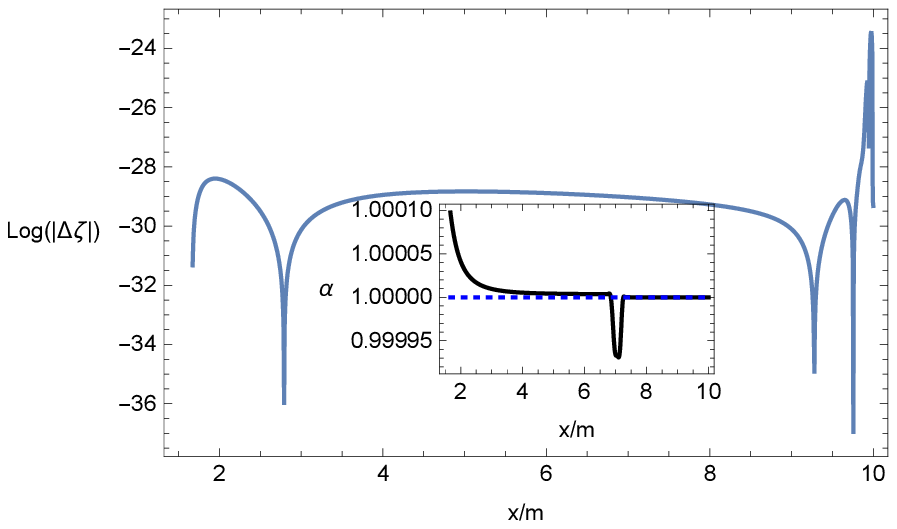}
\includegraphics[width=.314\textwidth]{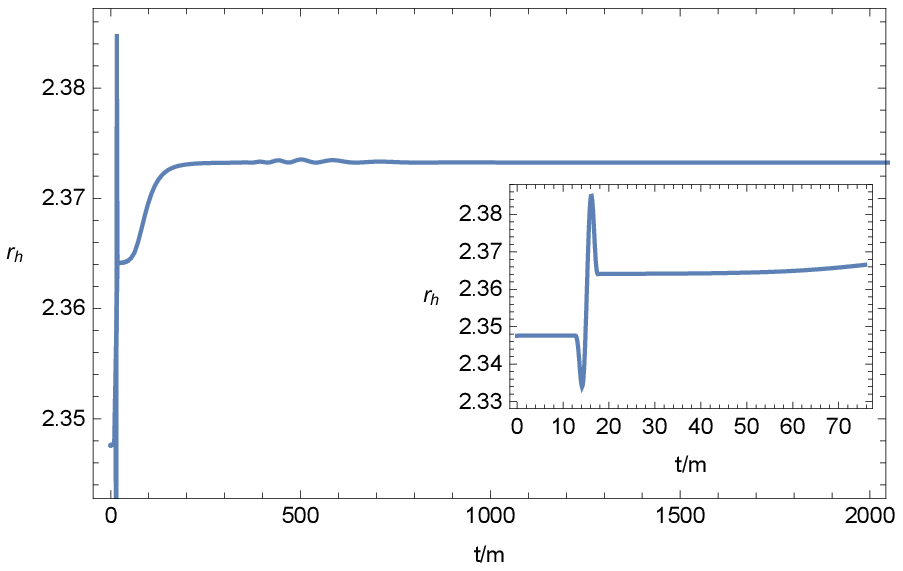}
\caption{\label{s-b-zeta}
The temporal evolutions of the relevant metric functions.
Left: the deviation $\Delta \zeta$ of the metric function $\zeta$ from that of the initial HBH metric, evaluated as a function of $x/m$ for different time instants.
Middle: the logarithmic deviation $\ln\left|\Delta \zeta\right|$ of the late-time asymptotic metric function $\zeta$ from that of a static BBH of the same mass, evaluated as a function of $x/m$. 
The solid black curve shown in the inset gives the initial profile of $\alpha$, while the blue dashed one corresponds to its value at $t/m=2180$.
Right: the radius of black hole horizon as a function of $t/m$.}
\end{figure} 

\begin{figure}[tbp]
\centering 
\includegraphics[width=.32\textwidth]{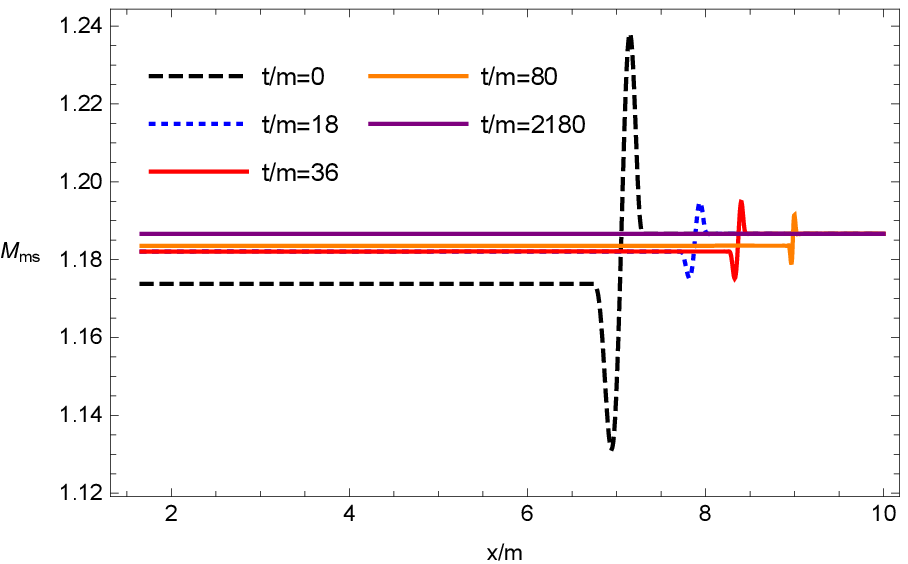}
\includegraphics[width=.32\textwidth]{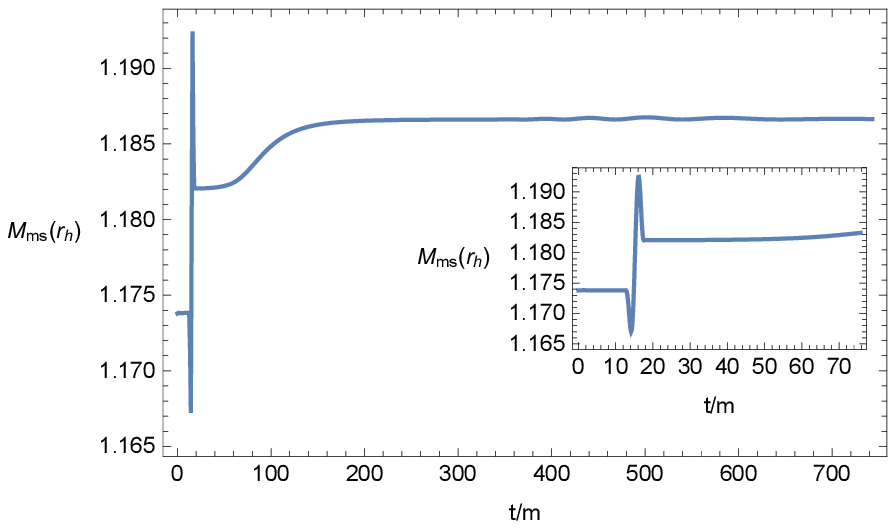}
\includegraphics[width=.33\textwidth]{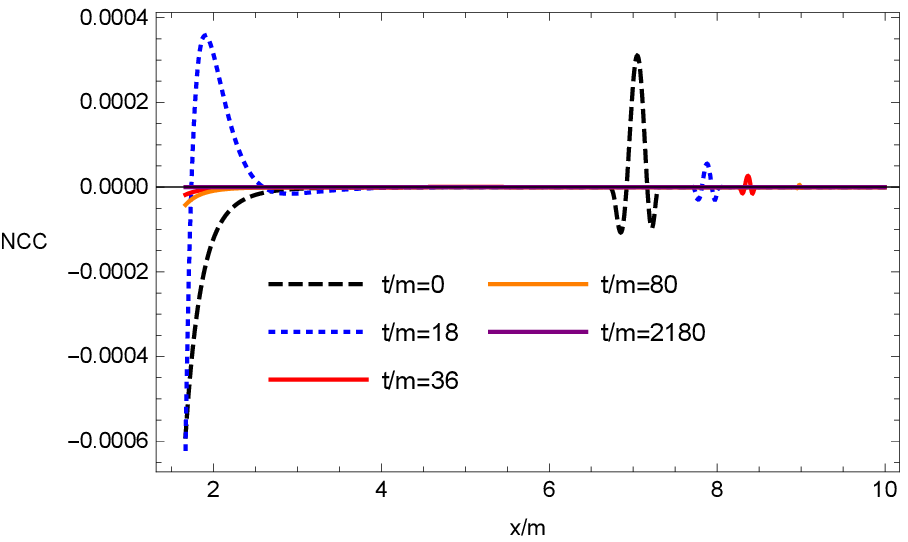}
\caption{\label{mass-deg} 
The evolution of MS mass $M_\mathrm{MS}$ and the scalar $f=R_{\mu\nu}l^{\mu}l^{\nu}$.
Left: the MS mass as a function of $x/m$, evaluated at different time instants.
Middle: the MS mass as a function of $t/m$ evaluated at the horizon.
Right: the scalar $f = R_{\mu\nu} l^\mu l^\nu$ as a function of $x/m$, evaluated at different time instants.
}
\end{figure}

The temporal evolutions of relevant metric quantities due to the energy injection are given in Fig.~\ref{s-b-zeta}.
In the left panel, we show the deviation of the metric function $\zeta(t,r)$ from that of the initial HBH, $\zeta_b(r)$, evaluated as a function of $x/m$ for different instants of time.
The black solid curve essentially describes the deviation from $\zeta_b(r)$ due to the energy injection of the initial scalar pulse.
As time processes, the pulse decays as the metric function $\zeta(t,r)$ becomes smoother.
At late times, It eventually converges to a given form where $\zeta(t,r)$ monotonously grows for smaller $x$.
The above asymptotic form, numerically evaluated at a sufficiently large instant $t/m=2180$, is then compared against the corresponding BBH of the same mass, as presented in the middle panel for the metric functions $\zeta$ and $\alpha$.
It is observed that the two profiles mostly coincide, where the difference is shown to be less than $e^{-24}$, from which one concludes that the final stage of the evolution is identical to a Schwarzschild BBH.
The right panel gives the black hole horizon $r_h$ versus time.
At early times, the size of the black hole horizon experiences some oscillation, and then it saturates at around $t/m\simeq200$.
To elaborate on the nonmonotonic evolution of the apparent horizon, we proceed to investigate the MS mass.
Fig.~\ref{mass-deg} explores the evolutions of MS mass $M_\mathrm{MS}$ and the scalar $f=R_{\mu\nu}l^{\mu}l^{\nu}$ as functions of the radial coordinate and time.
From the left panel, one observes that the initial pulse in the MS mass owing to the scalar perturbations propagates dissipatively toward the infinity.
As the MS mass evaluated at the horizon indicates the mass of the black hole, it is studied in more detail in the middle panel.
One finds that the MS mass at the horizon also oscillates for the time interval $12<t/m<18$, after which the MS mass converges to the value of the resulting BBH.
It is noted that the above time interval coincides with the arrival of the scalar pulse.
Therefore, such a nonmonotonic behavior of black hole mass might be interpreted as the scalar field extracting energy from the black hole, as the scalar pulse destabilizes the metric.
As discussed in the case of the route (b), this corresponds to the region where the energy density of the scalar field is negative.
Moreover, one observes that the NCC is transiently violated in two spacetime regions from the right panel.
As the scalar pulse destabilizes the metric near the horizon, NCC is temporarily violated.
Also, small regions of violation are observed, coinciding with the process when the initial scalar perturbations propagate outwardly to the spatial infinity.

\begin{figure}[tbp]
\centering 
\includegraphics[width=.323\textwidth]{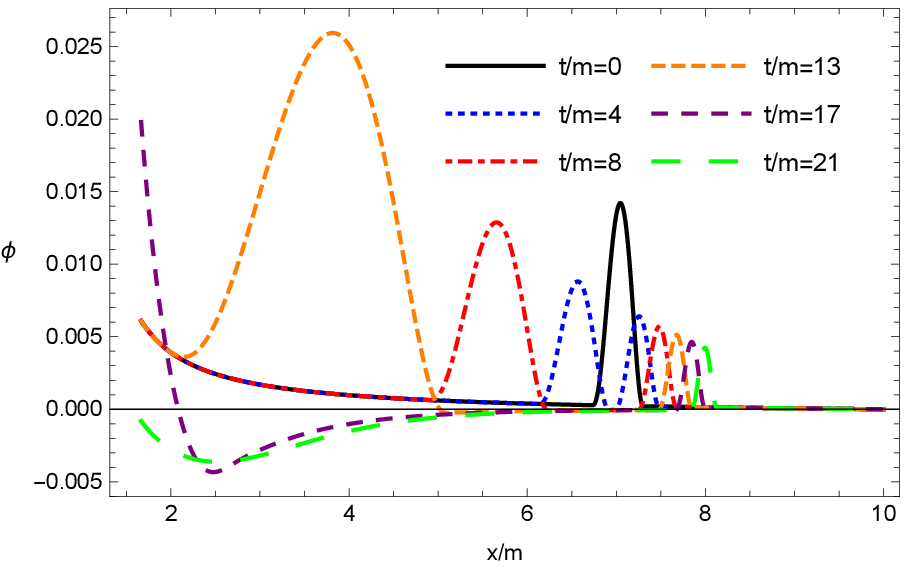}
\includegraphics[width=.32\textwidth]{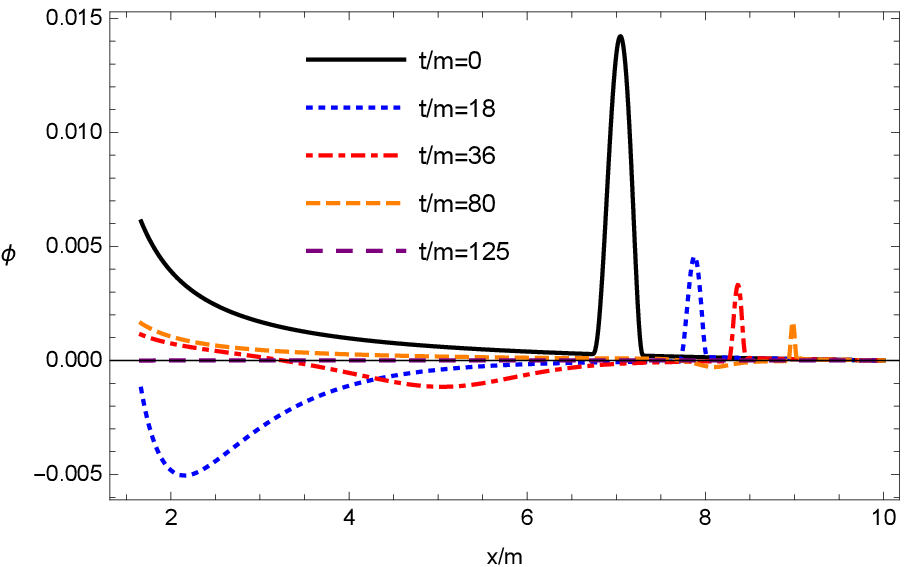}
\includegraphics[width=.316\textwidth]{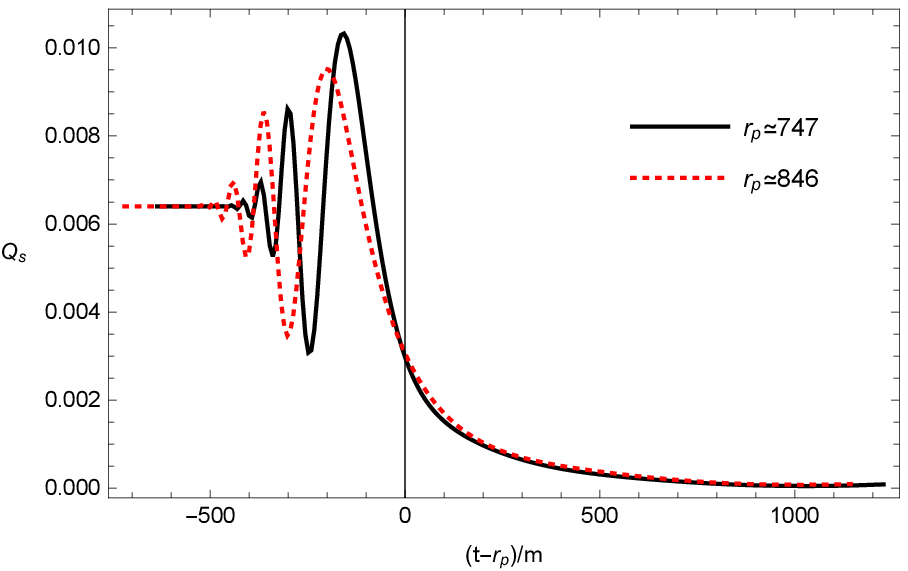}
\caption{\label{s-b-phi}
The evolution of the scalar field.
Left and Middle: the profiles of the scalar field, calculated at different time instants.
Bottom-left: the strength of the scalar field, evaluated at the horizon.
Right: the scalar charge $Q_s$ as a function of $(t-r_p)/m$, evaluated at $r_p= 747$ and $846$, respectively.}
\end{figure}

The results of the evolution of the scalar field are presented in Fig.~\ref{s-b-phi}.
The left and middle panels show the evolution of the scalar profile as a function of the radial coordinate at early and late times.
The black solid curve gives scalar field at $t=0$, which consists of two parts, namely, $\phi=\phi_b+\phi_p$.
As time increases, part of the scalar field propagates towards the horizon while the other part travels outward.
As shown in the upper-left panel, in the vicinity of the horizon, the distribution of the scalar field essentially remains the same until the arrival of the scalar pulse at about $t/m=13$.
Moreover, as the scalar pulse reaches the horizon, it destabilizes the spacetime configuration, as the strength of the scalar field raises temporarily. 
After a period of transient oscillations, during which the black hole partly absorbs the scalar field, the evolution leads to a significant suppression. 
Eventually, the scalar field vanishes identically from the entire spatial domain, as indicated by the upper-right panel.

The temporal evolution of the scalar field discussed above can also be seen by studying the scalar field evaluated at the black hole horizon as a function of $t/m$.
In the vicinity of the horizon, one finds that the scalar field first experiences a sharp rise at the arrival of the ingoing scalar pulse, which triggers the instability of the spacetime.
As the black hole begins to absorb the scalar field, the magnitude of the oscillations $\phi_h$ quickly decreases and eventually vanishes.
Subsequently, the resulting black hole is deprived of the scalar hair.

The bottom-right panel, we show the scalar charge as a function of time for a given large radial coordinate.
It is observed that the scalar charge also oscillates in the wake of the arrival of the ingoing scalar pulse.
Subsequently, it decays and eventually vanishes, agreeing with the above discussions.

\section{Concluding remarks}\label{conclusion}

In this work, we numerically investigated the far-from-equilibrium evolutions through which the scalar hair might be generated or deprived for the black holes in a gravitational theory where the Gauss-Bonnet and Ricci invariants are coupled to a scalar field.
In particular, we have explored dynamics of scalar field and spacetime in four distinct scenarios as indicated in Fig.~\ref{sketch}, disclosing the process of the scalar hair condensing onto the bald black hole to make the black hole become hairy, and how a hairy black hole loses its hair transiting to a bald one.
According to route (a), given an initial BBH with a mass beyond the threshold value, an injection of a scalar pulse may trigger a dynamic process through which the bald black hole swallows a fraction of the scalar perturbations and evolves into a heavier BBH.
Route (b) simulates the dynamical process of spontaneous scalarization.
In other words, an initial BBH with a mass below the threshold can be destabilized and evolve due to the interaction between the scalar perturbations.  The black hole can eventually acquire non-trivial scalar hair onto  the horizon.
We observed that the MS mass does not increase monotonously as a function of the radial coordinate, which is primarily related to the region where the quantity $f = R_{\mu\nu} l^\mu l^\nu$ becomes negative as the NCC is violated.
The MS mass evaluated at the black hole horizon increases in time monotonously, indicating that the black hole absorbs part of the scalar perturbations while the black hole radius increases.
As is illustrated by route (c), for an initial HBH with a mass below the threshold value, an injection of a scalar pulse can cause the black hole swallows part of the scalar perturbations and evolves into a heavier black hole with less scalar charge.
For route (d), by injection of strong enough energy into the gravitational system, the MS mass of the spacetime rises above the threshold while the black hole swallows its scalar hair.
As a result, the hair of the initial HBH is deprived, and the system evolves into a heavier stable BBH.
The present study revealed the rich underlying physics regarding the far-from-equilibrium evolution of gravitational systems.
This further exhibits that instead of the linear analysis, nonlinear dynamics is effective and powerful to disclose the final fate of the evolution, which crucially depends
on the initial spacetime configuration and the strength of the perturbations~\cite{Zhang:2021nnn}.

The appearance of ellipticity in the model seems to imply some unphysical degree of freedom. 
For instance, a ghost might be referred to when the system is either characterized by unbound energy, such as Ostrogradsky instability in higher-order theories, or gauge-dependent quanta, such as QCD ghost.
In practice, for the former case, one can explicitly remove the particular degree of freedom, while these ghosts are not malicious for the latter.
For partial differential equations, ellipticity is primarily indicated by the complex-valued characteristics.
As the coefficients of the equation in question are functions of spacetime coordinates, specific spacetime regions might be elliptic, which is notably coordiante independent.
In those regions, the system's dynamics cannot be posed as a well-defined Cauchy initial value problem, and one lost the notion of ``time direction''.
The corresponding treatment is also twofold.
As proposed in~\cite{Ripley:2019aqj}, the first possible solution is to eliminate such region from the underlying system in consideration.
In other words, one effectively removes the unphysical dynamics and the associated degree of freedom from the equations of motion.
The second possibility is that the ellipticity is benign. 
As for the present study, it is always hidden inside the horizon.
As a result, the latter does not lead to any observable implication.

In our scheme, the initial scalar pulse possesses a positive amount of energy, and numerically, the ADM mass of the resultant black hole spacetime is consistently found to increase.
As a result, all the arrows in Fig.~\ref{sketch} are directed to the right. 
This leads to the following question: even though the event horizon area always grows, is it possible that the mass of the scalar hair radiates away so fast that the total mass of the resulting static HBH spacetime decreases?
If the above scenario stands true, the arrow might be directed to the left, as the system involves in time.
Even though this was not observed in our numerical approach, from the theoretical perspective, the question is relevant in its own right.
Also, in the calculations, we have fixed the coupling constants $A$ and $B$ to simulate the far-from-equilibrium evolutions.
It would be rather interesting to study further how these coupling constants influence the dynamical process.
Besides, for the potential $V(\phi)$, we have chosen a specific form of the coupling function $W(\phi)$. 
It is not clear how the specific form of the potential would affect the underlying physics.
In practice, the hyperbolicity of the equations of motion has not posed a severe problem due to the specific metric parameters.
It will also be interesting to investigate further the parameter space, for which the system of equations of motion of the relevant theory dynamically loses the hyperbolicity.

\begin{acknowledgments} 
We thank the anonymous referee for the insightful questions and constructive comments.
This research is supported by the National Key R\&D Program of China under Grant No. $2020YFC2$ $201400$, the Major Program of the National Natural Science Foundation of China under Grant No. 11690021, and the National Natural Science Foundation of China under Grant Nos. 11505066, 11975235, 12005077 
and Guangdong Basic and Applied Basic Research Foundation under Grant No. 2021A1515012374. 
We also gratefully acknowledge the financial support from
Funda\c{c}\~ao de Amparo \`a Pesquisa do Estado de S\~ao Paulo (FAPESP), 
Funda\c{c}\~ao de Amparo \`a Pesquisa do Estado do Rio de Janeiro (FAPERJ), 
Conselho Nacional de Desenvolvimento Cient\'{\i}fico e Tecnol\'ogico (CNPq), 
Coordena\c{c}\~ao de Aperfei\c{c}oamento de Pessoal de N\'ivel Superior (CAPES).
Some of our calculations were performed using the tensor-algebra bundle xAct~\cite{xact}.

\end{acknowledgments}

\bibliographystyle{JHEP}
\bibliography{evolution}
\end{document}